\documentclass[11pt,twoside]{article}
\usepackage{asp2010}
\resetcounters
\aspvolume{***} 
\aspcpryear{2012} 
\aspvoltitle{Four Decades of Research on Massive Stars} 
\aspvolauthor{Laurent Drissen, Nicole St-Louis, Carmelle Robert, and
Anthony F. J. Moffat, eds.} 

\begin{document}

\title{Two Remarkable Spectroscopic Categories of Young O Stars from the
VLT-FLAMES Tarantula Survey}

\author{Nolan R.\ Walborn,$^1$ Hugues Sana,$^2$ William D.\ Taylor,$^3$ 
Sergio Sim\'on-D\'{\i}az,$^4$ and Christopher J.\ Evans$^5$}

\affil{$^1$Space Telescope Science Institute\\
3700 San Martin Drive, Baltimore, Maryland 21218, USA}

\affil{$^2$Astronomical Institute Anton Pannekoek, University of Amsterdam\\ 
Kruislaan 403, 1098 SJ, Amsterdam, The Netherlands}

\affil{$^3$Scottish Universities Physics Alliance, Institute for Astronomy, 
University of Edinburgh, Royal Observatory Edinburgh\\
Blackford Hill, Edinburgh, EH9 3HJ, UK}

\affil{$^4$Instituto de Astrof\'{\i}sica de Canarias\\ 
38200 La Laguna, Tenerife, Spain; and\\ 
Departamento de Astrof\'{\i}sica, Universidad de La Laguna\\
38205 La Laguna, Tenerife, Spain}

\affil{$^5$UK Astronomy Technology Centre, Royal Observatory Edinburgh\\
Blackford Hill, Edinburgh, EH9 3HJ, UK}

\begin{abstract}
The spectral and spatial characteristics of two special categories of O stars 
found in the VFTS dataset are presented.  One of them comprises very rapid 
rotators, including several more extreme than any previously known.  These 
objects are distributed around the peripheries of the main 30~Doradus clusters, 
suggesting a runaway nature for which their radial velocities already provide 
preliminary supporting evidence.  The other category consists of a large number 
of Vz stars, previously hypothesized on spectroscopic grounds to be on or 
very near the ZAMS.  Their distribution is the inverse of that of the rapid 
rotators: the Vz are strongly concentrated to the ionizing clusters, plus a 
newly recognized band of recent and current star formation to the north, which 
provides strong circumstantial evidence for their extreme youth.
\end{abstract}

\section{Introduction}
The VLT-FLAMES Tarantula Survey (VFTS; Evans et~al.\ 2011) has produced an
unprecedented spectroscopic dataset for the stellar content of the
30~Doradus Nebula in the Large Magellanic Cloud, the largest in the Local
Group.  About 800 OB stars have been observed with a resolving power of
$\sim$10$^4$ at the Very Large Telescope of the European Southern Observatory
(program 182.D-0222).  It is reasonable to expect that such an advance over
previous material will entail both observational and theoretical progress
on the formation and evolution of massive stars.  Indeed, some early,
unexpected empirical developments, resulting from the spectral classification
of 167 O stars with no detected radial-velocity variations, are briefly 
described here; they will be amplified shortly by Walborn et~al.\ (2012, 
in prep.), while the optimized classification atlas generated for this work 
will be presented by Sana et~al.\ (2012, in prep.).

\section{Rapid Rotators}
In this subsample, 17 O main-sequence stars have very high projected
rotational velocities in the range of 300--600~km~s$^{-1}$; the largest
values previously known were just over 400~km~s$^{-1}$ (Howarth \& Smith
2001; Walborn et~al.\ 2011).  An additional 7 rapid rotators have giant
luminosity classes indicated, although the actual luminosities of such
extreme objects may require further investigation.  The spectra of 6
luminosity class V stars are displayed in Figure~1, labeled with their
detailed spectral types and preliminary $v\sin i$ measurements derived by SSD 
using a Fourier transform technique.  (Definitive values will be derived by 
O.~Ram\'{\i}rez-Agudelo et~al.\ 2012 and C.~Sab\'{\i}n-Sanjuli\'an et~al.\ 2012, 
in prep.).  VFTS~285 is the current record holder.  Remarkably, none of 
these spectra displays Balmer central stellar emission, even at H$\alpha$,
indicating a possible absence of disks. A related case probably with a
disk is discussed by Dufton et~al.\ (2011).

\begin{figure}[!ht]
\centerline{\includegraphics[width=6.0in,angle=270]{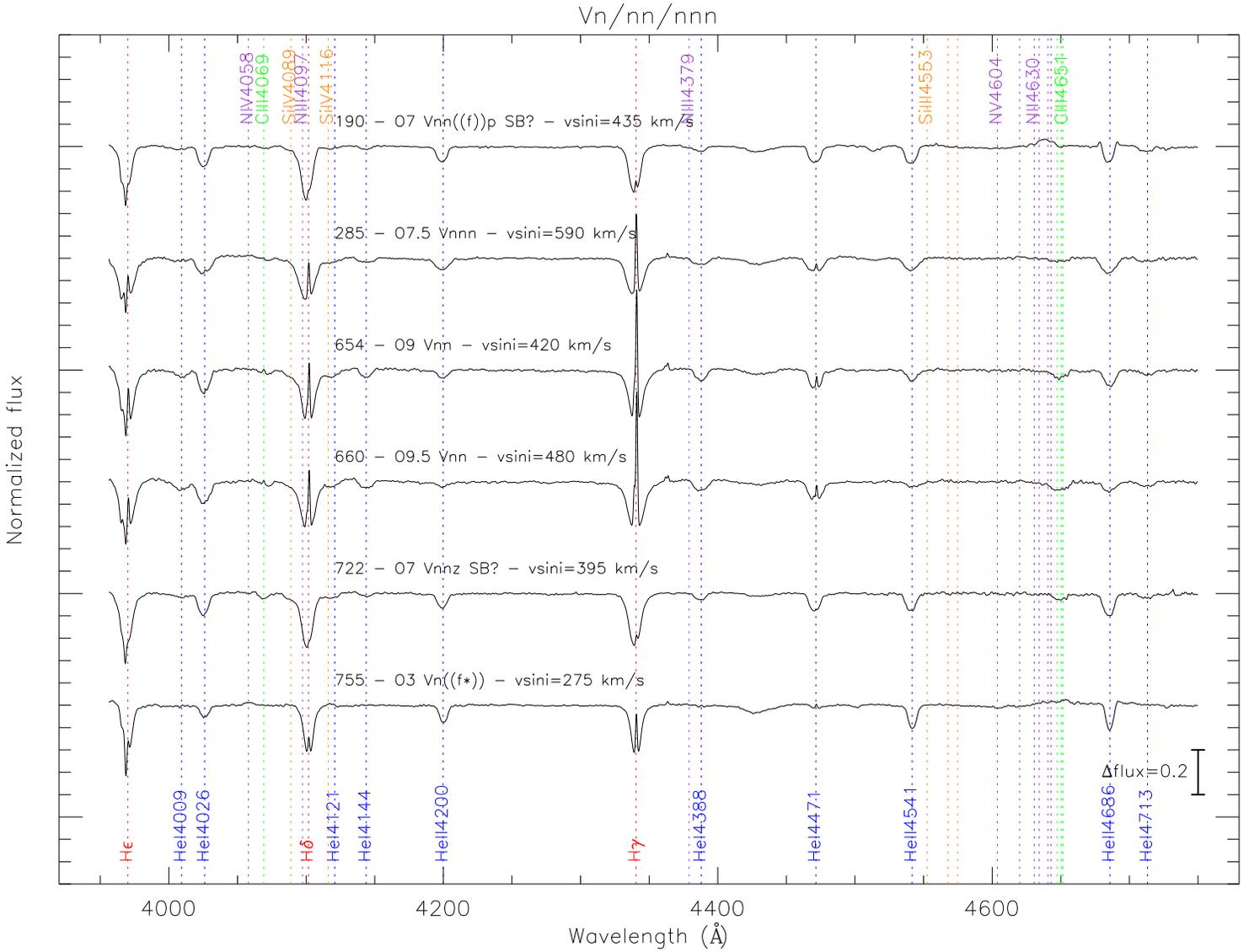}}
\caption{Rectified spectral intensity plots of extremely rapid rotators.
The VFTS catalogue numbers are followed by the spectral types and $v\sin i$
measurements.  VFTS 190 has an Onfp spectrum with emission wings at He~II
$\lambda$4686 (Walborn et~al.\ 2010a).  The central Balmer emission is nebular; 
``SB?'' signifies a relative displacement of the stellar absorption lines, 
although radial-velocity variations have not been detected in the available 
data.} 
\label{fig1}
\end{figure}

The spatial distribution of the main-sequence rapid rotators is shown in 
Figure~2.  It is seen that all but 4 of them are located about the peripheries 
of the clusters NGC~2060 and 2070.  This circumstance immediately suggests a
possible runaway nature of the class. The VFTS radial velocities of these stars 
already provide supporting evidence for that interpretation, containing a high 
fraction of large values compared to the rest of the sample (Sana et~al., 
these proceedings; et~al.\ 2012, in prep.).  A current {\it Hubble Space 
Telescope\/} imaging program by D.~Lennon et~al.\ may determine their proper 
motions, thus enabling a full kinematical analysis.  We emphasize that a 
population of high-mass, extreme-rotator runaways from a massive young cluster 
is a new phenomenon, which may be related to theoretical dynamical predictions 
and to the origin of gamma-ray bursts (Dale \& Davies 2006; Allison et~al.\ 
2010).  A possibly related situation for the Galactic cluster Westerlund~2 has 
been presented by Roman-Lopes, Barb\'a, \& Morrell (2011).

\begin{figure}[!ht]
\centerline{\includegraphics[width=4.0in]{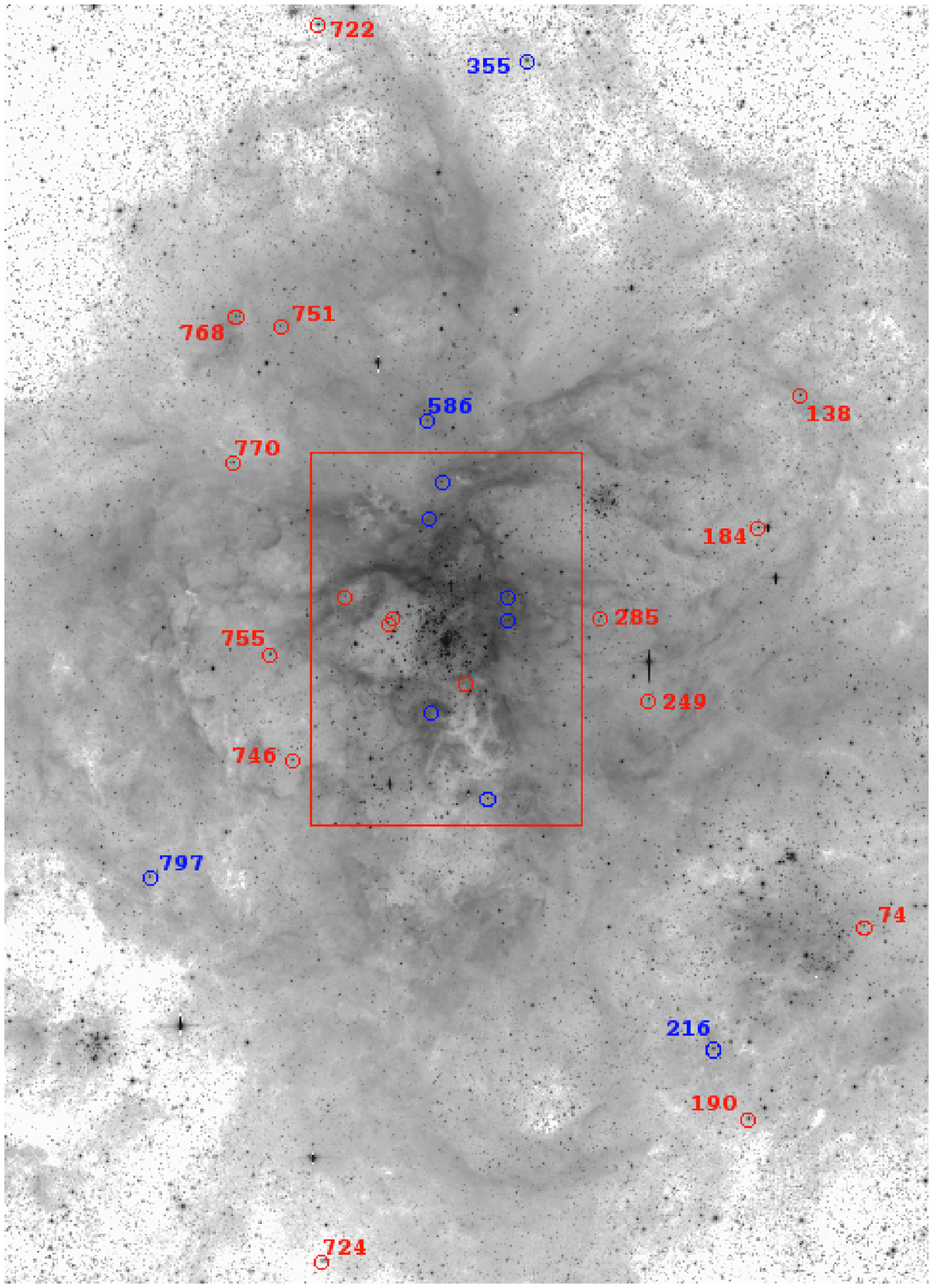}}
\caption{Distribution of the rapid rotators (VFTS numbers in red), on a 
logarithmic MPG/ESO/WFI $V$ image of 30~Doradus; VFTS 751 and 755 are separated
by 284\arcsec\ in declination, or 71~pc in projection.  North is up and east to 
the left.  NGC~2060 is at the SW and NGC~2070 in the center.  The unannotated 
red circles in the core are VFTS 465, 660, 654, and 706.  The blue symbols 
correspond to O~V((fc)) spectra (see Fig.~3 caption), which do not show any 
preferred locations.} 
\label{fig2}
\end{figure}

\section{Zero-Age Main-Sequence Stars}
A class of O-type spectra with empirical evidence of subluminosity is
described by Walborn (2009).  Briefly, they display He~II $\lambda4686$
absorption stronger than any other He~II or He~I line.  Since emission
in that particular He~II line is a progressive luminosity effect (the Of 
phenomenon), these spectra, assigned luminosity class Vz, have been
hypothesized to represent the inverse effect, i.e. lower luminosity, 
higher gravity, and/or lower mass-loss rate than for normal class V.  
Although such spectra have systematically been found in young regions, 
we were surprised to encounter no fewer than 35 definite plus another 10 
possible examples in the VFTS subsample discussed here.  The spectra of a 
few of them are reproduced in Figure~3 (along with examples of some other 
special class V categories mentioned in the caption).  The physical 
parameters of this category have not been investigated previously, but 
they will be in the extensive VFTS sample.

\begin{figure}[!ht]
\centerline{\includegraphics[width=6.0in,angle=270]{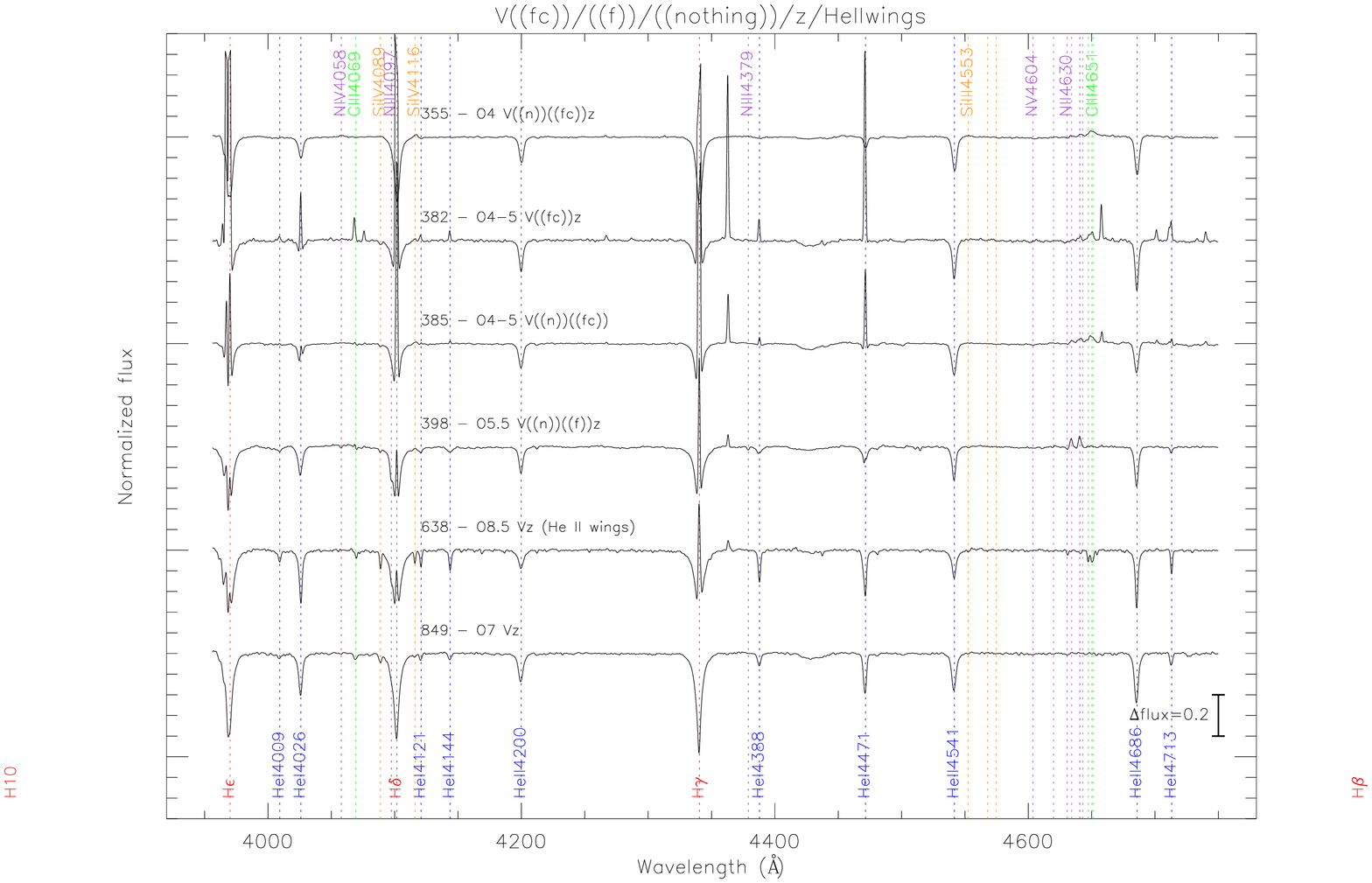}}
\caption{Rectified spectral intensity plots of Vz and some other subcategories 
of O-type main-sequence spectra, including V((fc)) with C~III $\lambda$4650 
emission lines comparable to N~III $\lambda$4640 (Walborn et~al.\ 2010b), 
He~II wings stronger than He~I (VFTS 638), and no N~III or C~III emission 
features whatever despite very high S/N (VFTS 849).  Note that Vz spectra
can be either ((f)) or ((fc)), or neither.}
\label{fig3}
\end{figure}

The distribution of the definite Vz objects is shown in Figure~4.  It is
essentially the inverse of the rapid rotator distribution, with a strong
concentration toward the main clusters NGC~2060 and 2070.  In addition,
several of the Vz stars lie in an approximately east-west band at the
northern extreme of the field.  The ``isolated'' WN stars R144, R146, and
R147 (Feast, Thackeray, \& Wesselink 1960) also lie in this band, as does 
one of the most luminous IR YSOs in the {\it Spitzer} images.  Thus, this band 
evidently represents another, previously unrecognized region of recent and 
current star formation in 30~Dor, consistent with the presence of Vz stars 
there.  This distribution provides strong circumstantial evidence for extreme 
youth of the Vz class, to be further investigated by quantitative analysis 
of the VFTS spectroscopic data.

\begin{figure}[!ht]
\plotone{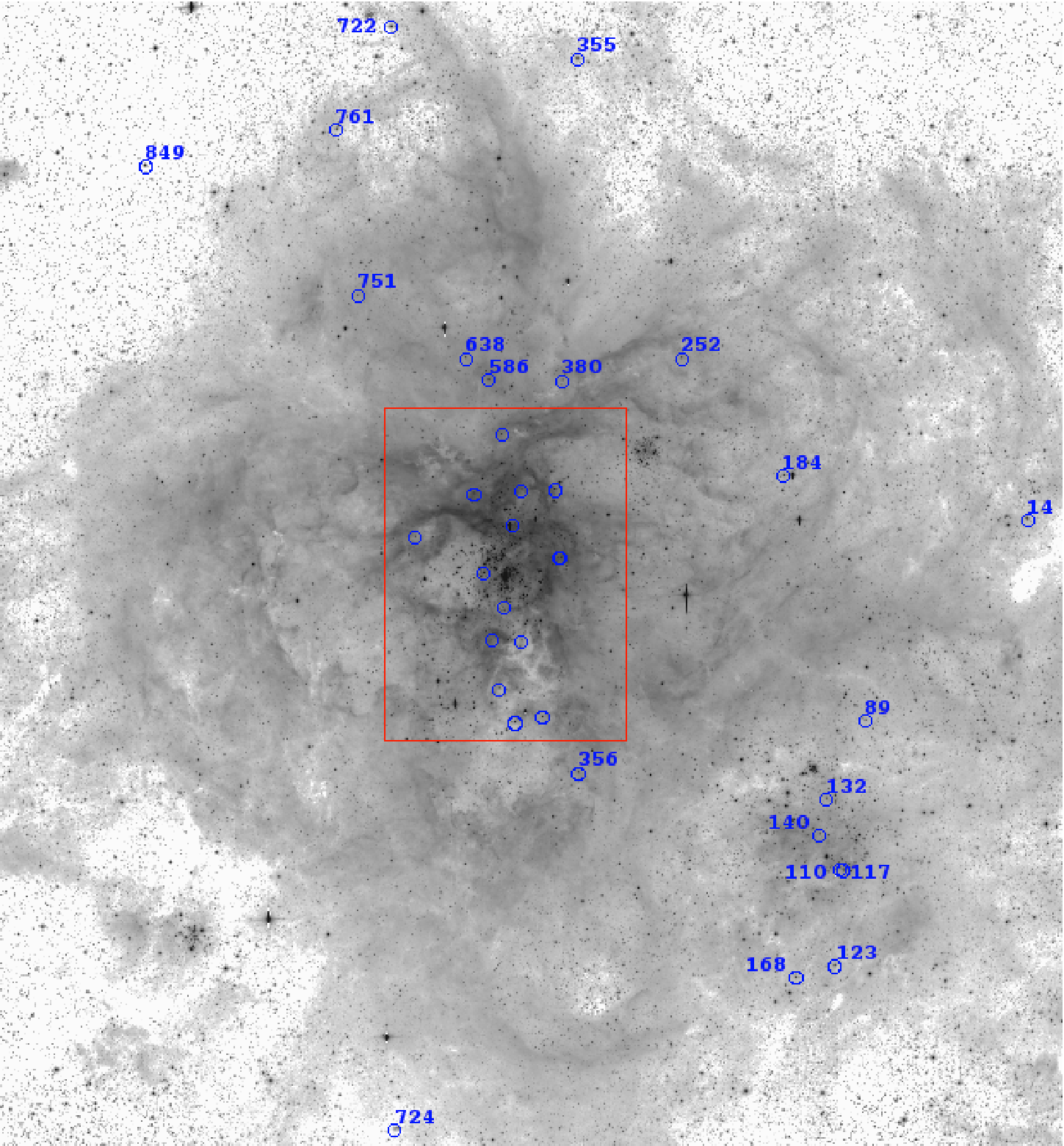}
\caption{Distribution of the Vz objects; image as in Fig.~2.  The numerous 
stars in the central cluster will be identified by Walborn et~al.\ (2012),
where the complete spectral classifications will also be given.}
\label{fig4}
\end{figure}

\acknowledgements NRW's travel to Lac Taureau was supported by NASA
through grant GO-12179.01 from STScI, which is operated by AURA, Inc.\ 
under contract NAS5-26555.  Thanks to A.~de~Koter for comments that
improved the clarity of the presentation.


\end{document}